# OBSERVATIONAL AND THEORETICAL CONSTRAINTS ON SINGULAR DARK MATTER HALOS

RICARDO A. FLORES*

Departmento de Física, Facultad de Ciencias Físicas y Matemáticas

Universidad de Chile, Casilla 487-3, Santiago, CHILE

and

JOEL R. PRIMACK

Santa Cruz Institute for Particle Physics, University of California, Santa Cruz, CA 95064, U.S.A.

## Abstract

The distribution of dark matter around galactic or cluster halos has usually been assumed to be approximately isothermal with a non-zero core radius, which is expected to be of the order of the size of the visible matter distribution. Recently, the possibility has been raised that dark matter halos might be singular in the sense that the dark matter density $\rho$ could increase monotonically with radius $r$ down to a very small distance from the center of galaxies or clusters. Such central cusps in the dark matter density could lead to a high flux of gamma rays from WIMP dark matter annihilation. Here we analyze two possibilities that have been discussed in the literature, $\rho \propto r^{-n}$ with $n \approx 1$ or $2$, and point out that such density profiles are excluded by gravitational lensing analyses on cluster scales and by the rotation curves of gas-rich, halo-dominated dwarf spirals on small scales. We also point out that if spiral galaxies form by gas infall inside dark matter halos, as they are expected to do in any hierarchical clustering model, such profiles almost always lead to falling rotation curves after infall, contrary to observations.

---

* On leave from Physics Department, University of Missouri, Saint Louis, MO 63121

The structure of dark matter (DM) halos around galaxies and clusters is not very well known. The best information about it comes from the data on the rotation of spiral galaxies, which restrict the density of DM to decrease roughly with the inverse square of the distance from the center, $\rho \propto r^{-2}$, from distances comparable to their optical radii, out to the largest distances at which neutral hydrogen is detected (see e.g. Casertano & van Gorkom 1991 [CvG], Flores *et al.* 1993 [FPBF], and references therein). Such is precisely the density profile of an isothermal sphere at radii much larger than its core radius, and it has traditionally been assumed that DM halos are approximately isothermal. This is theoretically favored since one would expect the collapse of DM halos from the overall expansion of the universe to be fairly chaotic, and thus the conditions of violent relaxation to apply (Lynden-Bell 1967, Shu 1978). In such a case, the equilibrium configuration is an isothermal sphere.

However, the chaotic changes in the gravitational potential during the collapse of DM halos occur for a limited time, and violent relaxation will therefore be incomplete. The collapse of DM halos, and their equilibrium configuration, has been studied via N–body simulations. Cosmological N–body simulations indeed found DM halos with approximate $\rho \propto r^{-2}$ profiles (Frenk *et al.* 1985; Quinn, Salmon, & Zurek 1986). Unfortunately, these simulations lacked the resolution to uncover a finite core radius or find deviations from this profile in the inner parts. Recent simulations (Dubinski & Carlberg 1991; Warren *et al.* 1992), however, have probed the structure of DM halos down to very small scales. Dubinski & Carlberg (1991) have simulated the collapse of isolated, large protogalaxies with tidal-field boundary conditions and found that the equilibrium structure of Cold Dark Matter (CDM) halos was well approximated by a Hernquist (1990) profile, for which $\rho \propto r^{-1}$ in the inner parts, down to the smallest scale resolved, $r \approx 1$ kpc. Using a different approach to approximate the non-linear evolution of DM halos, on the other hand, Berezinsky, Gurevich, & Zybin (1992) have argued that the DM distribution would have $\rho \propto r^{-n}$, with $n \approx 1.8$ down to scales as small as a fraction of a pc. They claim that their argument should apply from small galaxy scales up to cluster scales.

Such cusps in the dark matter distribution would have very interesting implications for particle DM searches. For example, it might then be possible to detect gamma rays from annihilations of very heavy DM particles in the center of our galaxy (Urban *et al.* 1992), and present limits on radio and gamma ray emission from the galactic center would then significantly constrain the mass of DM particles such as neutralinos (Berezinsky, Gurevich, & Zybin 1992).

In this paper we study whether observations can constrain or rule out the possibility that DM halos could have such monotonically increasing density profiles down to very small scales. We find that such "singular" halo profiles are inconsistent with (a) the rotation curve data of well-studied, gas-rich dwarf spirals such as DDO154 (Carignan & Freeman 1988, Carignan & Beaulieu 1989) and DDO168 (Broeils 1992), and (b) the measured distortion of distant galaxy shapes by the gravitational lensing of two clusters (Tyson, Valdes, & Wenk 1990 [TVW]); and that (c) if spiral galaxies form by gas infall inside DM halos, as they are expected to do in any hierarchical clustering model, their rotation curves would almost always fall with distance outside their optical radii in such a case, contrary to observations. We discuss each of these observational problems for singular halos in turn, and summarize our conclusions afterwards.

### (a) Dwarf Spiral Galaxies

The unusual properties of the dwarf spiral galaxy DDO154 (Carignan & Freeman 1988, Carignan & Beaulieu 1989) make it an ideal laboratory to test the distribution of DM in galaxies, at least on small scales. The luminous components of DDO154 are neutral hydrogen gas, visible through 21cm emission, and stars. With a gas–mass–to–blue–light ratio $M_{\rm HI}/L_B \approx 5$, which makes it a very gas–rich system, DDO154 offers the advantage that the mass of the dominant luminous component, the gas, is quite insensitive to the uncertain stellar–mass–to–light ratio. The inferred contribution of the luminous components to the circular velocity reveals a system dominated by its DM down to very small distances from the center; within the largest distance at which the HI is detected, more than 90% of the mass is invisible. Thus, the DM distribution is very well constrained in this system.

In Figure 1 (a) we show the inferred contribution of the DM to the circular velocity as a function of the distance $r$ from the center of DDO154. We have used the parameters of Carignan & Beaulieu (1989) for the two luminous components to extract the DM contribution from the data; thus we model the stellar component as a thin exponential disk of mass and scale length $M_* = 5 \times 10^7$ M$_\odot$ and $b_* = 0.5$ kpc. The HI surface density profile of DDO154 is well approximated for $r \gtrsim b_{HI}$ by an exponential profile, of scale length $b_{HI} = 1.65$ kpc; however, it deviates significantly from this profile for $r \lesssim b_{HI}$. Therefore, we model the HI component as a thin disk with surface density profile $\sigma_{HI}(r)$ fit to the data: $\sigma_{HI}(r) = (5.81 {\rm M}_\odot {\rm pc}^{-2}) \exp -(r/2b_{HI})^2$ for $r \leq 2b_{HI}$, and $(15.78 {\rm M}_\odot {\rm pc}^{-2}) \exp -(r/b_{HI})$ for $r \geq 2b_{HI}$. Finally, we scale the HI surface density by 1.3 in order to take He into account.

We calculate the contribution of the stellar and gaseous components to the circular velocity (see *e.g.* Binney & Tremaine 1987) and extract it in quadrature from the measured circular velocity to obtain the points plotted in the figure.

The dashed line in Figure 1 (a) shows the curve expected if $\rho \propto r^{-1.8}$, as argued by Berezinsky, Gurevich, & Zybin (1992), eye–fit to the large velocity data. We also show the curve expected for a Hernquist density profile (dotted line), again eye–fit to the large velocity data. The disagreement with the data at small radii for both profiles is quite clear, and based on the figure we conclude that they are excluded by these data. Note, however, that the CDM simulations of Dubinski & Carlberg (1991) might not apply to small galaxies like DDO154 because the fluctuation spectrum would be a little different on such scales. The data are consistent, on the other hand, with an isothermal profile of non–zero core radius; the solid line is the linear rise expected inside the core radius, where $\rho \approx$ constant, eye–fit to the small velocity data.

The rotation curve of DDO154 is by no means peculiar. Other dwarf spirals have measured rotation curves that are also incompatible with the singular halos we have discussed, although the inferred DM contribution to the rotational velocity is more sensitive to the assumed stellar disk mass because they are not equally gas rich. In Figure 1 (b) we show the constraints on the halo rotation curve of DDO168 (Broeils 1992), for which $M_{HI}/L_B \approx 1$. Here we have used a very small stellar mass–to–light ratio ($M_*/L_B = 0.1$) suggested by the best-fit mass model of Broeils (1992) to extract the constraints, thus the inferred DM rotation curve shown is almost as high as it can be at small radii. We conclude from this figure that the singular halo profiles are also incompatible with these data. We have also analyzed the rotation curve data of NGC3109 (Jobin & Carignan 1990) with similar results.

It is perhaps worth emphasizing here that two potentially important sources of systematic errors in the observations discussed above, asymmetric drift and disk warping, are taken into account in obtaining the circular velocity data in all three cases (Carignan & Beaulieu 1989, Jobin & Carignan 1990, Broeils 1992). The smallness of their residual velocity fields attest to the quality of the smoothed velocity fields fit to the raw data.

*(b) Cluster Gravitational Lensing Data*

Much data has accumulated over the last few years on lensing effects by clusters on background sources (for a recent review on lensing see e.g. Blandford & Narayan 1992).

For example, giant arcs have been studied in detail in order to study the distribution of DM in clusters (see e.g. Mellier, Fort, & Kneib 1993; Miralda-Escudé 1993; and references therein). The total mass of the lensing cluster can be fairly well determined and it agrees with virial estimates, but the distribution of the mass is not very well constrained. Small core radii $r_{core} \gtrsim 50$ kpc are infered in general, but a recent study (Bartelmann & Weiss 1993) suggests that clusters with somewhat more extended cores may not only be compatible with the lensing data but also with the cluster X–ray data and CDM N–body simulations.

TVW measured the shape distortion of very distant blue galaxies caused by the gravitational lensing of their light by two foreground clusters. The magnitude of the effect depends on the clusters' mass, and the masses determined also agreed with virial estimates. Their analysis provided for the first time a determination of the distribution of the DM inside clusters. For example, simply associating all the clusters' mass with the brightest galaxies is not allowed by their data, and they showed that a *finite* core radius is required if the DM distribution is fit to an isothermal sphere.

The distortion measured by TVW gives us a simple way to constrain the density profiles considered here because it does not depend on details of the images. The relationship between the measured distortion and the mass distribution has been rigorously studied by Kaiser & Squires (1993) [KS]. In order to compare the relative magnitude of the distortion that would be generated by clusters with the various profiles we are considering, we have calculated the TVW distortion generated by such profiles following the analysis of KS. The distortion is given by

$$D(r) = \frac{1}{2\pi} \int T(k)\Sigma(k)J_0(kr)kdk , \qquad (1)$$

where $T(k) = k^{-2}(2(1 - J_0(kr_{cut})) - kr_{cut}J_1(kr_{cut}))$, $\Sigma(k)$ is the Fourier transform of the normalized surface density profile, the $J$'s are Bessel functions, and $r_{cut} \approx 300$ kpc is the cutoff radius introduced by TVW.

In Figure 2 we show the distortion, Equation (1), for the various profiles, normalized to unity at small radii. The TVW data for Abell 1689 (from Fig. 3 of KS) are shown by the points. The DM core radius of the isothermal distribution shown is $r_{core} = 50$ kpc. We show two cases of a Hernquist profile (for which $\rho \propto r^{-1}(r+r_H)^{-3}$), $r_H = r_{core}$ and $r_H = 10r_{core}$, for the following reasons. Dubinski & Carlberg (1991) found $r_H \approx 30$ kpc in their simulation of a $10^{12}$ M$_\odot$ halo, which is of the order of the core radius needed for an isothermal profile that fits the rotation curve of a large spiral inmersed in such a halo. Thus, we show the case $r_H = r_{core}$ in Figure 2. On the other hand if one wanted a Hernquist profile and an isothermal profile to have similar outer rotation curves out to $r_{cut}$, a large $r_H$ is needed. Thus, we show a representative case with $r_H = 10 r_{core}$. It can be seen that all the singular profiles but the Hernquist profile with $r_H = 10 r_{core}$ lead to a distortion that continues to rise well inside $r_{cut}$, which is not seen in the TVW data. It will be interesting to see whether new data and analyses agree.

### (c) Gas Infall and Large–Spiral Rotation Curves

Galaxies are expected to form within DM halos in any hierarchical clustering (bottom–up) cosmology. Their luminous matter is a certain fraction $F$ of the total mass in the volume where the gas has cooled, and $F$ is observationally constrained to lie in the range $0.05 \lesssim F \lesssim 0.2$ (Blumenthal *et al.* 1984). The contraction of the gas, as it cools within the DM halo, significantly concentrates the DM around the galaxy, even if the fraction of $F$ gas is as small as $\sim 5-10\%$ (BFFP; Carlberg, Lake, & Norman 1986). The amount of radial collapse of the gas in the case of spiral galaxies is expected to be controlled by the protogalaxy's initial angular momentum, and one can then determine theoretical spiral rotation curves in the combined potential of their exponential disks and contracted halos (FPFB, and references therein). DM halos must be fairly diffuse before gas infall, or else spiral rotation curves always decline outside their optical radii (FPFB), contrary to observations (see e.g. CvG and references therein).

In Figure 3 we plot a measure of the rotation curve slope expected for spiral galaxies if they had formed inside DM halos that had the Hernquist or the $r^{-1.8}$ profile before gas infall, shown as a function of the dimensionless protogalactic angular momentum $\lambda$. For the Hernquist profile we have taken the ratio of the scale radius $r_H$ and the maximum radius of gas infall $R$ (see FPBF) to be $r_H/R = 1/2$, since $r_H \approx 30$ kpc and $R \gtrsim 50\sqrt{F/0.01}$ kpc in the quasistatic cooling limit; smaller values of this ratio yield smaller slopes. HI data have been used by CvG and Broeils (1992) to study the slope of spiral rotation curves, mostly at radial distances outside their optical radii. The logarithmic slope of galaxies with maximum circular velocities in excess of 100 km/sec, defined as the slope of the log-log rotation curve plot $\chi^2$-fit to a straight line, covers the entire range $-0.2 \lesssim \text{logslope} \lesssim 0.1$ (CvG, Broeils 1992; smaller galaxies have even more positive slopes, but we do not necessarily expect our calculations to apply to them because a significant gas loss would be expected as a result of supernova driven winds — see e.g. Dekel & Silk 1986). For a value of the gas fraction at the lower end of the allowed range, $F = 0.05$, we see that neither the Hernquist nor the $r^{-1.8}$ density profiles would result in positive logarithmic slopes for $\lambda \leq 0.15$. But note that N–body simulations have shown that most protogalaxies ($\sim 80\%$) would be expected to have $0.02 \lesssim \lambda \lesssim 0.1$, and that protogalaxies with $\lambda \geq 0.15$ would be extremely rare (Barnes & Efstathiou 1987). As can be seen in the Figure, extremely small gas fractions $F \approx 0.01$ would be required to yield logarithmic slopes as large as 0.1. If $F$ were a universal constant or its range of values were scale independent, this is definitely excluded by observational constraints (Blumenthal *et al.* 1984). Small gas fractions only on large galaxy scales, $0.01 \leq F \leq 0.05$, could be compatible with these constraints if galaxies with logslope $\geq 0$ were very rare, but this is not the case: 6 out of 23 galaxies with maximum circular velocities in excess of 100 km/sec have logslope $\geq 0$. Thus, we conclude that this possibility is also excluded. However, the sample of galaxies with HI data is too small to establish confidence levels.

One potentially important limitation of the gas infall model we have used (BFFP, FPBF) is that it assumes that the gas cools within an already virialized DM halo, despite the fact that recent numerical simulations (Katz & Gunn 1991, Katz 1992) indicate that gas disks might form simultaneously with the virialization of the DM halo. However, BFFP showed that applying the infall model to the equilibrium configuration resulting from the dissipationless virialization of a DM halo produced a rotation curve that agreed quite well, at the large radii we have discussed, with the rotation curve determined from their dissipational N–body simulations, in which most of the dissipation does occur during the crunch after a protogalaxy stops expanding and collapses.

*Conclusions*

We have studied the possibility that DM halos could be coreless distributions with steeply increasing density of dark matter toward their centers. We have found that such DM density cusps in small galaxies and clusters are incompatible with observations, and that disk galaxy formation inside such halos yields disk rotation curves incompatible with observational constraints. Our work has several implications. (1) The DM halo of our galaxy is unlikely to have a central density large enough to yield a detectable gamma ray flux from annihilations of very heavy DM particles (see Urban *et al.* 1992); however, our results do not rule out the possibility that such enhancement would occur if the dark matter virialized and its distribution became isothermal in the presence of an already formed compact luminous component (see Ipser & Sikivie 1987). (2) Our results considerably weaken the mass limits on DM particles of Berezinsky, Gurevich, & Zybin 1992. Our results pose a constraint on DM clustering models. Gurevich & Zybin 1988 have argued that gravitational clustering would yield $r^{-1.8}$ halos of very small core radii, regardless of the power spectrum of density fluctuations. However, their analysis of the gravitational collapse of DM halos neglects the kinetic energy that is present as a result of the density fluctuations on small scales. Thus, we do not believe this to be a general prediction of the gravitational instability theory. (3) It is not clear to us whether the singular halos found by Dubinski & Carlberg are really a necessary consequence of CDM models; it could be that their result is an artifact due to approximations in their calculations, for example the way the angular momentum of their protogalaxy was generated. If not, then it is possible that the nature of the dark matter distribution at small radii can shed light on the nature of dark matter in galaxies and clusters. For example, inclusion in simulations of gas (e.g. Katz 1992) or a significant hot dark matter component (Davis, Sommers, & Schlegel 1992; Klypin *et al.* 1993) along with cold dark matter could affect the dark matter halo density profile. This is undoubtedly an important issue that must be addressed by self-consistent, high resolution cosmological simulations of gravitational collapse.

We thank S.N. Dutta, G. Blumenthal, J. Ellis, S.M. Faber, L. Hernquist, D. Spergel, and M. Way for helpful conversations and/or email. We especially thank N. Kaiser for helping us understand the TVW analysis, as in KS. We also thank the Aspen Center for Physics, where some of this work began, for its hospitality. This work has been partially supported by a fellowship from Fundación Andes (RF), and NSF grants.

# FIGURE CAPTIONS

**Figure 1** – DM contribution to the circular velocity about the center of (a) DDO154, and (b) DDO168, both as a function of distance from the center in units of the HI–disk scale length. The circular velocity plotted was obtained by subtraction from the data of the contribution due to the two luminous matter components (see text). The lines show the radial behavior expected if the dark matter had a Hernquist (dotted), $r^{-1.8}$ (dashed), or constant (solid) density profile.

**Figure 2** – Distortion $D(r)$, Equation (1), as a function of distance $r$ to the lens's center. The points show the TVW data for Abell 1689. The lines show the radial behavior expected if the dark matter had a Hernquist (dotted), $r^{-1.8}$ (dashed), or isothermal (solid, $r_{core} = 50$ kpc) density profile. Shown for comparison is an isothermal profile with $r_{core} = 0$ (dash–dotted). See text for explanations.

**Figure 3** – Logarithmic slope of the theoretical rotation curves expected for spiral galaxies that formed inside a DM halo that had a Hernquist (dotted) or $r^{-1.8}$ (dashed) density profile, shown as a function of the dimensionless protogalactic angular momentum $\lambda$ and for two values of the gas fraction $F$. The vertical line shows the observed range of values of the logarithmic slope of large galaxies.

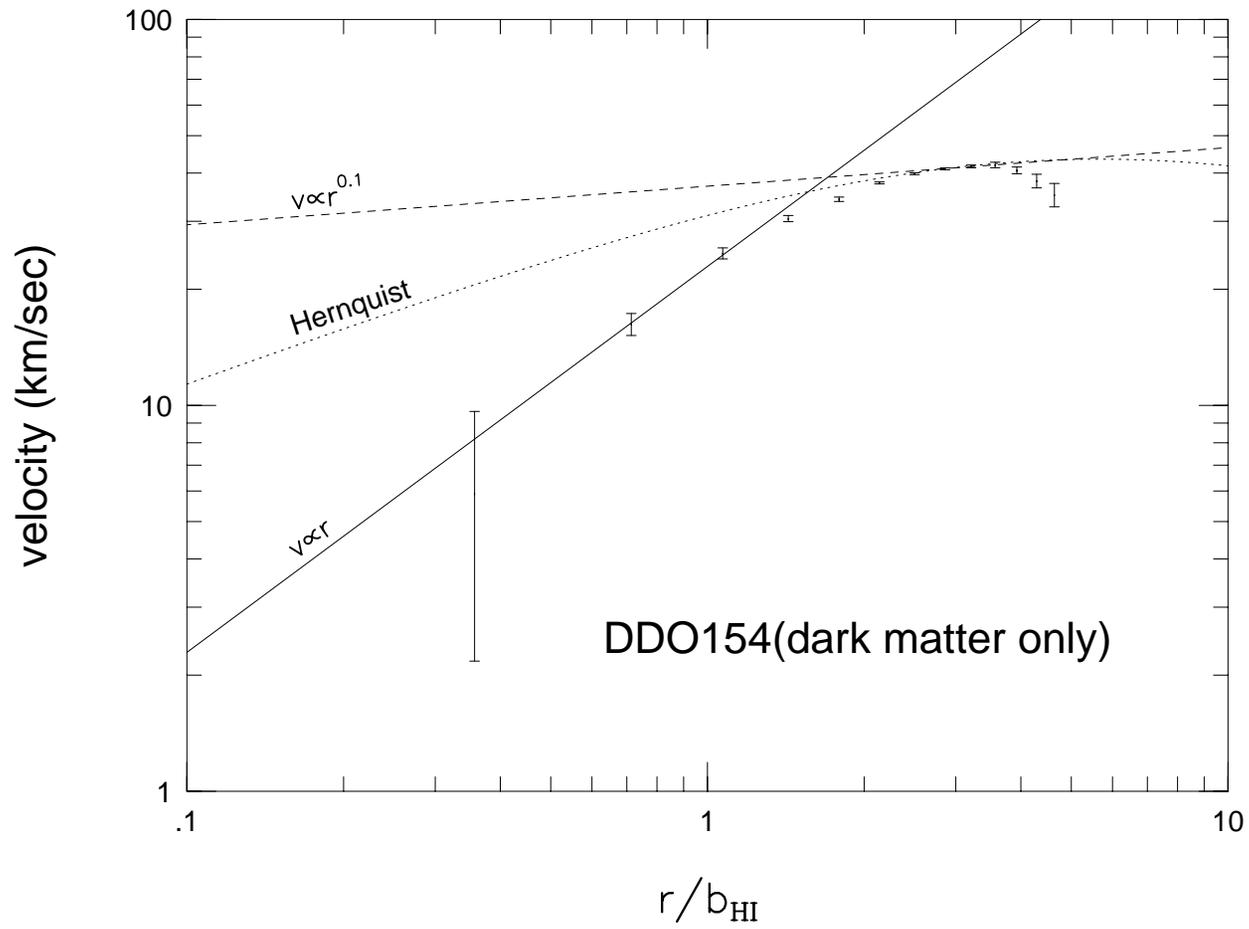

Figure 1(a)

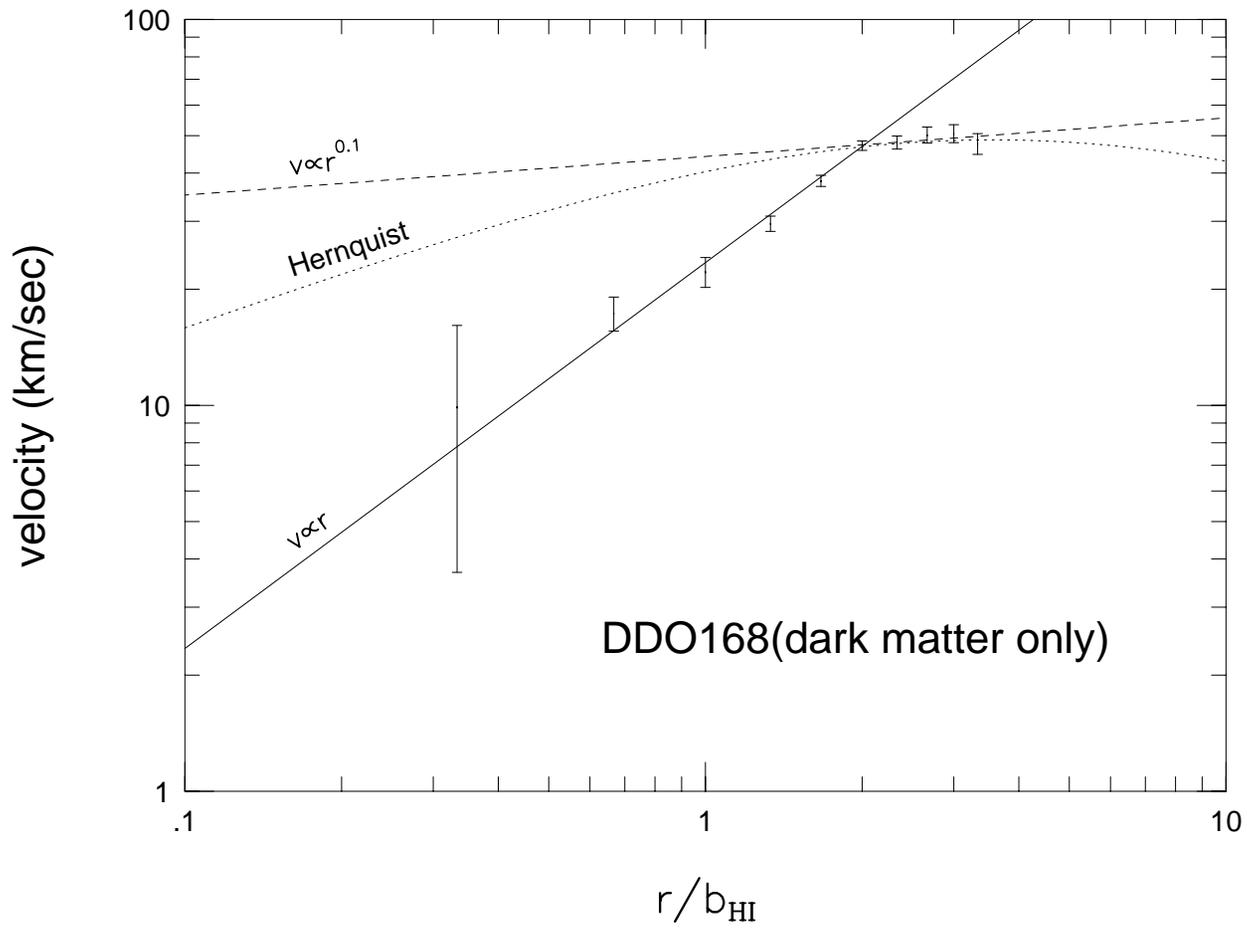

Figure 1(b)

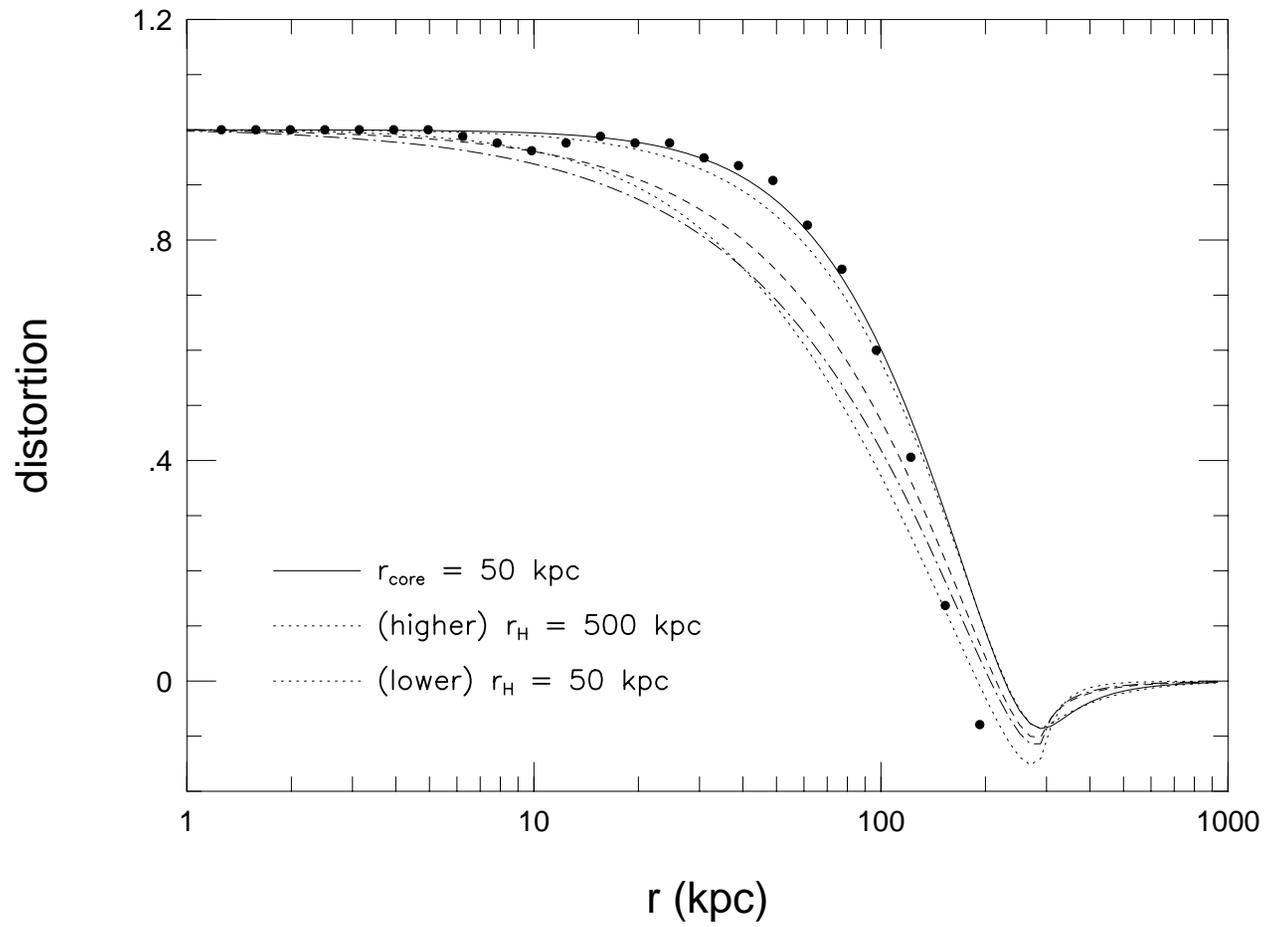

Figure 2

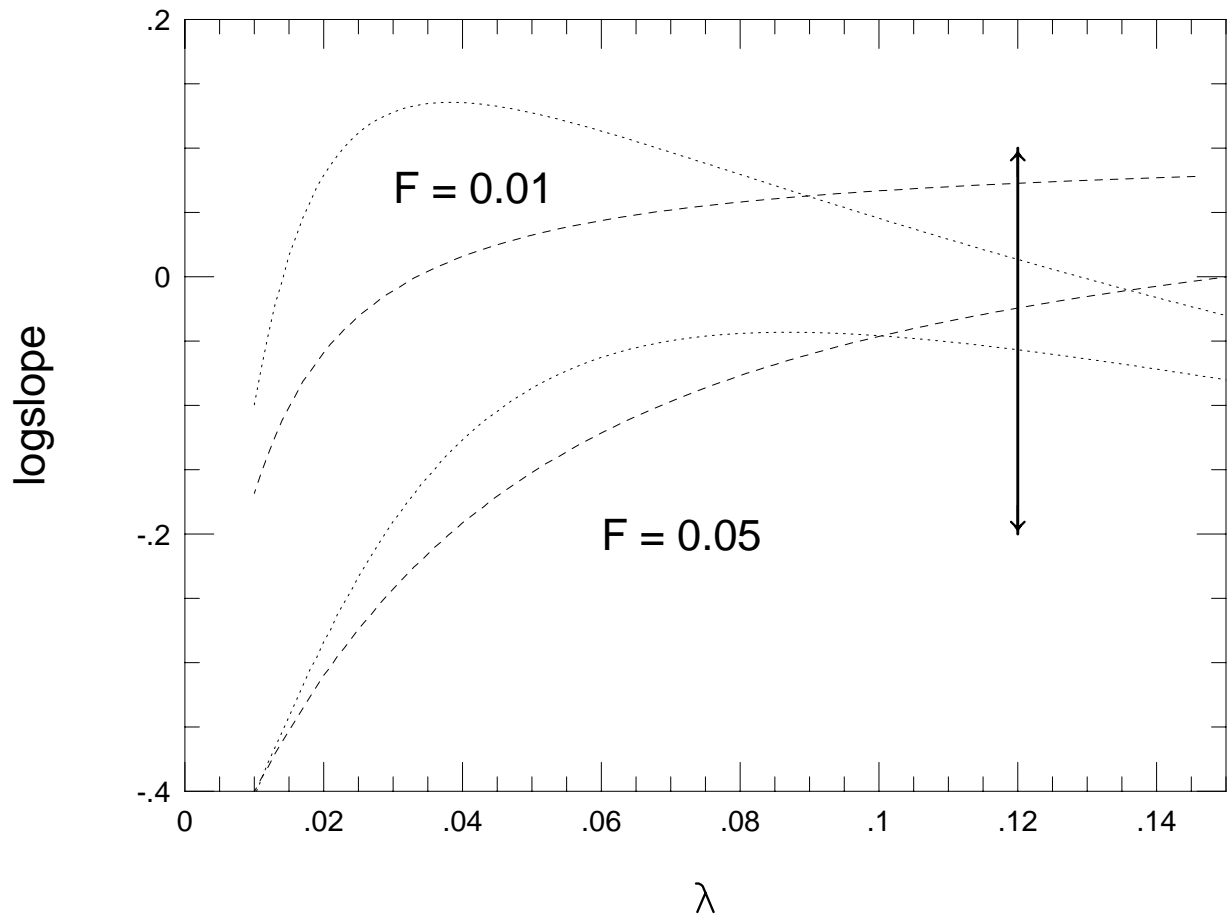

Figure 3